\documentclass[aps,prl,twocolumn,groupedaddress,showpacs]{revtex4}

\usepackage{epsfig,amsmath,amsbsy,amssymb}
\usepackage[T1]{fontenc}

\usepackage{graphicx}
\usepackage{bm}

\usepackage{ifthen}

\usepackage{color,colordvi}
\definecolor{Red}{rgb}{0.9,0.0,0.1}

\definecolor{Green}{rgb}{0,0.9,0.1}

\begin{document}

\newcommand{\lK}{\left(}
\newcommand{\rK}{\right)}

\title{Wrinkling of Random and Regular Semiflexible Polymer Networks}

\author{Pascal M\"uller}
\affiliation{Physics Department, TU Dortmund University, 
44221 Dortmund, Germany}
\author{Jan Kierfeld}
\affiliation{Physics Department, TU Dortmund University, 
44221 Dortmund, Germany}

\date{\today}
\begin{abstract}
We investigate wrinkling of two-dimensional 
random and triangular semiflexible 
polymer networks under shear. 
Both types of semiflexible networks 
exhibit wrinkling above a small  critical shear angle,
which scales with an exponent of the bending modulus between 
$1.9$ and $2.0$.
Random networks exhibit hysteresis at the wrinkling threshold. 
Wrinkling lowers the total elastic energy by up to $20\%$ and
strongly affects the elastic properties of all
semiflexible networks such as the crossover between bending and 
stretching dominated behavior.
In random  networks, we also find evidence for metastable 
wrinkled configurations. While 
the disordered microstructure of random networks affects the scaling 
behavior of wrinkle amplitudes, it has little 
effect on wrinkle wavelength. 
Therefore, wrinkles represent a robust, microstructure-independent 
assay of shear strain or elastic properties. 
\end{abstract}

\pacs{46.32.+x, 87.16.Ka,  87.16.dm, 46.70.De}
                             

\maketitle

Random networks of stiff fibers or polymers are important 
 model systems for biopolymer  meshworks in the 
cell cortex or the cytoplasm \cite{Kasza2007} or synthetic
nanofibrous materials \cite{Burger2006}.
In such a network structure, a random array of 
fibers or semiflexible polymers with an 
intrinsic bending rigidity is cross-linked. We will focus on the 
situation where cross-linking is permanent on the time scale of 
network deformation. Quasi-two-dimensional (2D) 
geometries are used to model sheetlike 
materials such as the cell cortex, the spectrin cytoskeleton 
of red blood cells or synthetic materials such as paper sheets. 
In fact, 
much of the theoretical and simulation 
work  on semiflexible polymer networks has been 
done on planar
networks \cite{Head2003,Wilhelm2003,Head2003b,Heussinger2006,
Onck2005,Broedersz2011}; this work has shown that
planar networks  exhibit unique elastic properties under shear
with a crossover from bending 
to stretching dominated elasticity, nonaffine deformation, and 
strain hardening.
More recently, similar properties could be found in
simulations of 
 three-dimensional bulk networks, which are suitable models for 
bulk biopolymer gels 
 \cite{Buxton2007,Huisman2007,Blundell2009,Huisman2011,Broedersz2011}.

Realistic 2D sheetlike 
materials are, however, embedded into three-dimensional space 
and  exhibit a buckling instability 
with deformations normal to the initial plane 
resulting in wrinkling even if applied 
stresses are strictly in plane. All existing works 
on 2D semiflexible networks neglect this issue.
We show that wrinkling  strongly modifies the 
elastic properties of a 2D sheetlike material under shear
in comparison to  strictly planar deformations.

Wrinkling of a 2D material 
can occur under different loading conditions:
Stretching with clamped boundaries  causes tensional wrinkles 
along the stretching direction \cite{Cerda2002,Cerda2003,Davidovitch2011},
and shearing gives rise to  wrinkles at a 
$45^{\circ}$-angle \cite{timoshenko,Wong2006}. 
In both cases, the material is under compressive stress perpendicular to 
the wrinkle orientation.
Here, we want to study wrinkling of
2D networks of semiflexible 
polymers under shear deformation in order to provide a more 
 realistic treatment of  sheet-like networks 
embedded into three-dimensional space, where
the effects of wrinkling are of practical relevance for applications 
of these materials. We will compare
regular and disordered network geometries to address the
 important question in the context of wrinkling theory 
to what extent the wrinkling instability and wrinkle properties 
depend on the 
microstructure of a material.
We will compare our findings for  2D semiflexible networks 
to the continuum linear elasticity  theory 
of wrinkling \cite{timoshenko,Wong2006}.

\begin{figure}
\begin{center}
\includegraphics[width=\linewidth]{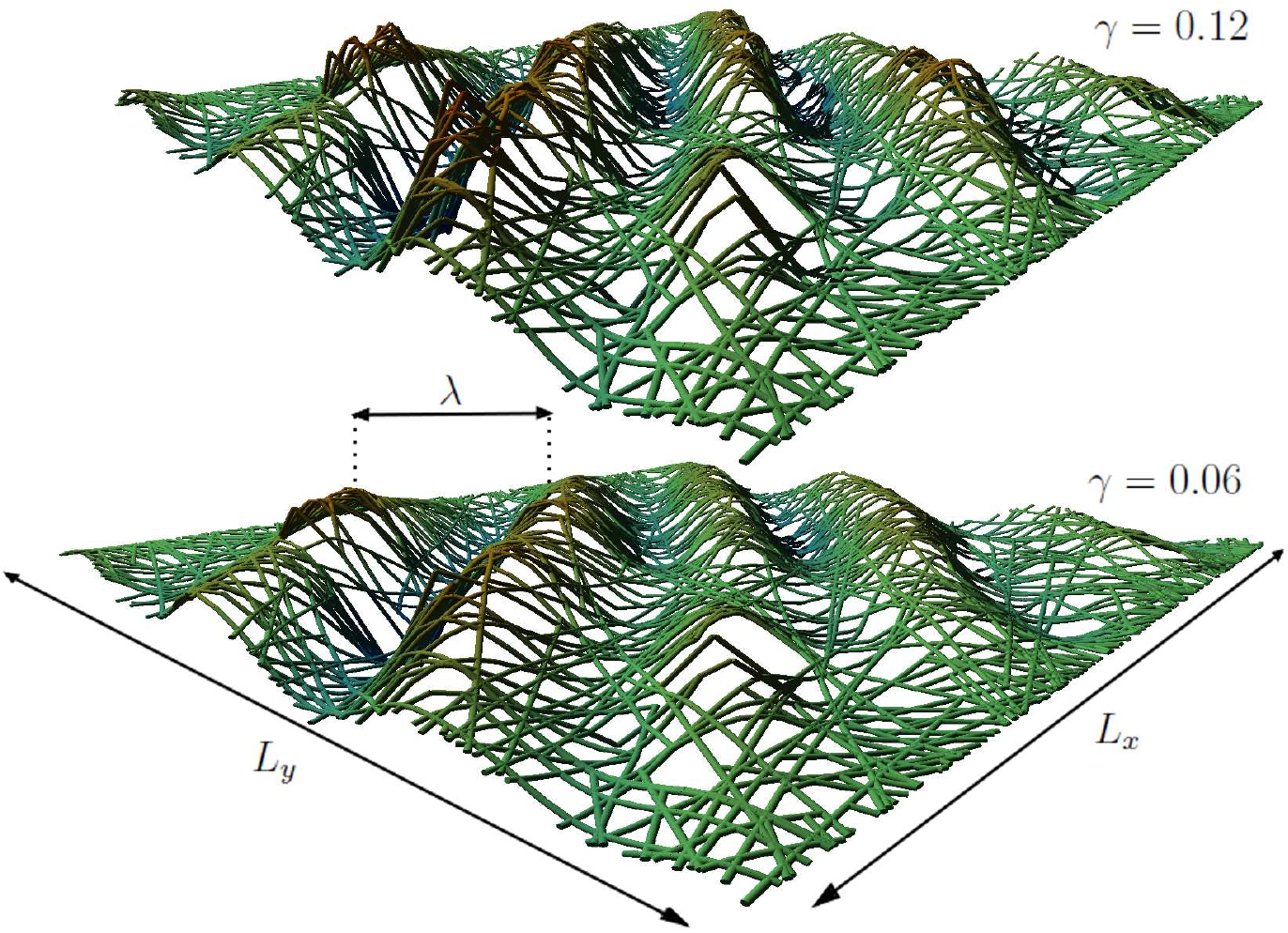}
\caption{
   Snapshots of a wrinkled random network ($l_b/L = 2.5 \times 10^{-3}$ 
   and $\eta = 54$) at  two different shear strains $\gamma=0.06$ and 
   $0.12$. The dimensionless displacement $z/\lambda$ in normal direction 
  is color coded. 
   Wrinkles form at an
  angle of $45^\circ$ to the shearing direction; the wrinkle amplitude 
   increases with the shear angle.}
\label{fig:wrex}
\end{center}
\end{figure}

\paragraph{Model and Simulation.}

Semiflexible polymer  networks are generated by placing straight rods 
of length $L$ into a 2D
rectangular simulation cell with dimensions $L_x\times L_y$ we use 
$L_x/L= L_y/L = 1.5$ for random networks). Each rod is
characterized by a  stretching modulus $\mu$ and a
bending modulus $\kappa$. We neglect thermal fluctuations and only
consider mechanical elasticity. Hence, the energy of a filament
consists of a stretching energy contribution $E^{(s)} = \int
 (\mu/2) [u'(s)]^2 ds$ and a bending energy
contribution $E^{(b)} = \int (\kappa/2) [\phi'(s)]^2 ds$. 
Here, $u'(s)$ is the local strain and
$\phi'(s)$ is the local curvature, both parametrized by the
contour length $s$. The elastic moduli can be used to define the
length scale  $l_b \equiv \sqrt{\kappa / \mu}$, 
which can  be interpreted
as the ``thickness'' of the network as we will argue below. 
The  number $l_b/L$ is our dimensionless measure for the 
relative bending rigidity of the rods. For F-actin of length
 $L=20\,{\rm \mu m}$, 
we typically find  $l_b/L \simeq 10^{-4}$;
throughout the Letter, we consider comparable or higher 
bending rigidities 
 $l_b/L = 1.4\times 10^{-4},...,2.5\times 10^{-3}$.

Filament intersections  are identified as cross-links. During
simulation, the cross-links are treated as permanent and freely
rotating. Filaments exceeding the simulation cell in the $y$ direction are
fixed to the $y$ boundaries. In the $x$ direction, we use periodic
boundary conditions. 
Depending on the depositing routine, two different
types of 2D networks are generated: (i) The rods are either placed
equidistantly at fixed orientations, resulting in a regular triangular
network
or (ii) added at random positions with random orientations, forming a
disordered, random network. 
In a triangular network (i), 
the filaments extend through the whole simulation cell, and the lattice
constant $l_c^{\vartriangle}$ is specified to determine the total
number of filaments. 
In a disordered network (ii), 
filaments are added until the
network density reaches a specified value. We calculate the dimensionless 
network
density as $\eta = L/l_c$. Here, $l_c$ is the mean distance
between neighboring cross-links, and  $\eta$ is equivalent to the
average number of cross-links per rod. 
Throughout the Letter we consider densities 
$\eta= 32,...,54$. 
For random networks, this is far above 
the rigidity percolation threshold, 
which is  around $\eta_p \simeq 6$ \cite{latva2001}. 
Based on Ref.\ \cite{Liu2007}, we estimate 
typical values for F-actin networks as $5 \lesssim \eta \lesssim 100$.
For regular networks, we 
define an analogous  density parameter $\eta' = L_x/l_c^{\vartriangle}$. 
In order to allow bending of
the segments between cross-links, the midpoints 
of each segment can be displaced. 
With this discretization, only the first bending mode
is considered, which is expected to dominate in the absence of thermal 
fluctuations \cite{Head2003}.

The elastic moduli of a regular triangular network can be calculated
analytically.  We find a  shear modulus $G^{\vartriangle} =
 \sqrt{3}\mu/4l_c^{\vartriangle}$ and  a  Young's
modulus $Y^{\vartriangle} = 8 G^{\vartriangle}/3$ resulting in  a Poisson
ratio  $\nu^{\vartriangle} = 1/3$ and a  
bending modulus $B^{\vartriangle} = 3G^{\vartriangle}\kappa/\mu$.
The densities of the random networks 
considered in this Letter are sufficiently 
high that 
the elastic moduli can be calculated by assuming affine 
deformations and a uniform  distribution of rod 
orientations \cite{Head2003, Head2003b}. This leads to a shear modulus 
$G^r = (\pi/16) (\mu/L) (\eta + 2/\eta - 3)$
and relations for the other moduli which
are  identical to regular  triangular networks: $Y^r = 8 G^r/3$,
$\nu^r = 1/3$, and $B^r = 3G^r \kappa/\mu$. Using  these
results, we can generate regular and random networks with the same
elastic moduli for comparison.
Both for triangular and random networks we have 
  $B/3G = \kappa/\mu = l_b^2$.
In shell elasticity, the ratio $B/3G$ 
is proportional to the square of the 
shell thickness and in elasticity theory
of  thin cylindrical rods  of diameter $d$, we have
$\kappa/\mu \sim d^2$ \cite{landau1986}, 
which  suggests identifying $l_b$ with a network 
``thickness''.

The degrees of freedom of the  simulation are the positions of 
cross-links and segment midpoints. Starting from an undeformed planar 
state, the
shear deformation $\gamma$ in the $x$ direction is increased by small
increments $\delta \gamma$. For each increment, we  perform an
affine deformation of all points,  fix the $y$ boundaries at their
new positions, and find the  configuration of minimal total energy
using  a conjugate gradient algorithm. To allow for 
wrinkling, we allow cross-link and midpoint positions to move in all 
three spatial dimensions. To avoid getting trapped in the metastable planar 
network configuration and enable wrinkle formation,
 the $z$ coordinates (normal to the 
initial network plane)  of cross-links and midpoints 
 are randomly perturbed before  minimization.
Without this perturbation, the networks remain planar under shear,
hence, they do  not exhibit wrinkling. This enables us to 
 simulate the same networks
with and without wrinkling and 
compare their total energies $E_{wr}$ (wrinkled) and $E_{pl}$ (planar).
Details of the simulation model are contained in 
the Supplemental Material
  \cite{EPAPS}.

\paragraph{Wrinkle formation.}

\begin{figure}
\begin{center}
\includegraphics[width=\linewidth]{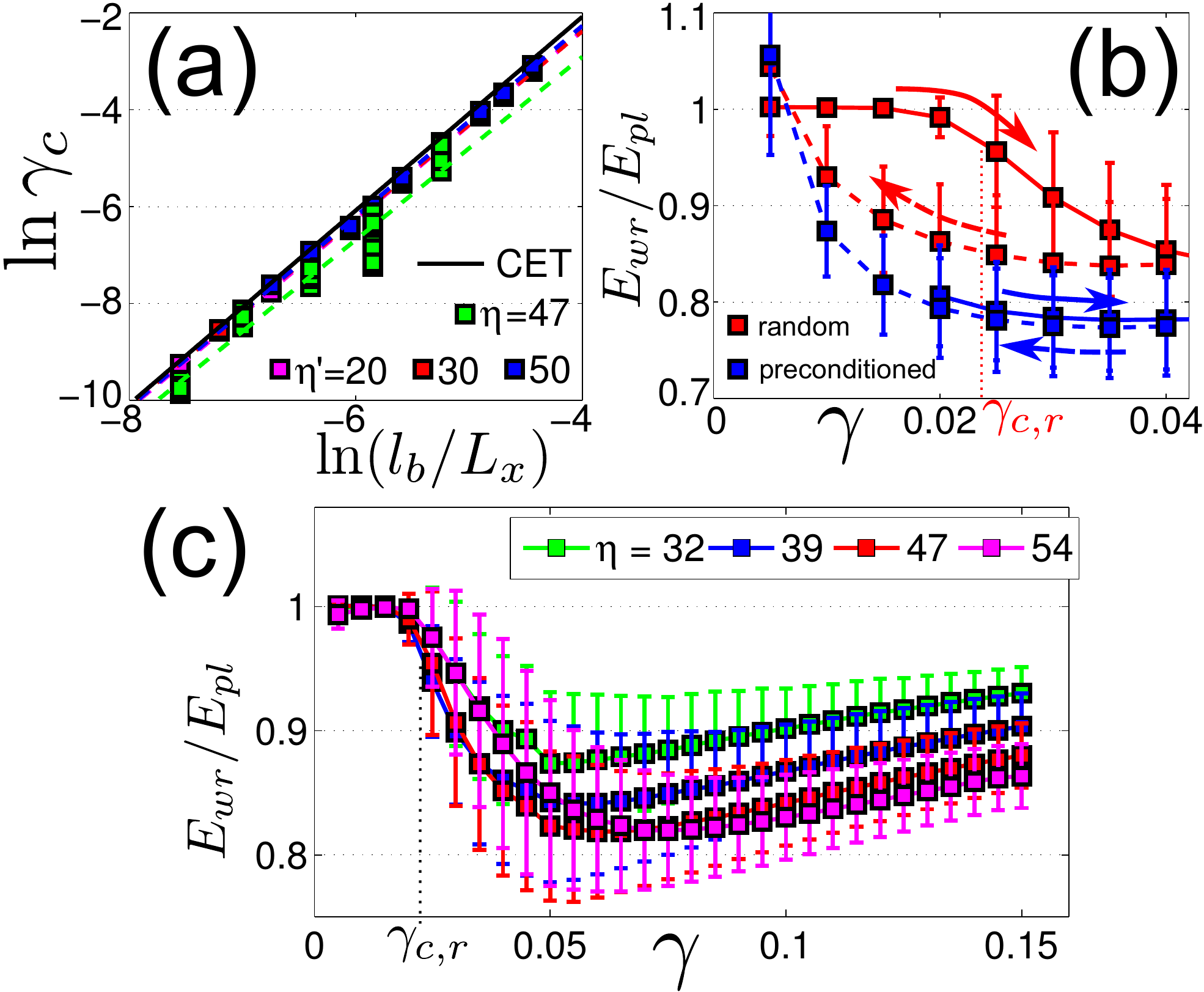}
\caption{
  (a) Critical shear angle $\gamma_c$  
   as a function of the rigidity (double logarithmic) 
    for  random  (densities $\eta=47$)  and regular networks 
   ($\eta'=20,\,30,\,50$) in comparison with linear continuum elasticity
   theory (CET).
  Lines  are  least square fits  $\gamma_c \propto l_b^{\alpha}$ (see text).
  (b) Total energy ratio $E_{wr}/E_{pl}$
  of wrinkled to planar networks as a function of shear angle $\gamma$ 
   for  random networks (red)   with rigidity
   $l_b/L = 2.5\times 10^{-3}$,  $\eta  =47$.
   We find hysteresis upon reversing the 
    deformation as indicated by the arrows.
    Preconditioning  random networks  with wrinkle patterns from regular 
   networks results in a lower energy (blue).
  (c) Total energy ratio $E_{wr}/E_{pl}$
 as a function of shear angle $\gamma$ 
   for  random  networks with  
    $l_b/L = 2.5\times 10^{-3}$ and $\eta = 32,\,39,\,47,\,54$.
      Data points in (b),(c) represent averages
    over ten  realizations, and the error bars indicate the
    statistical spread.}
\label{fig:gcEhy}
\end{center}
\end{figure}

For shear angles $\gamma$ exceeding a critical value
$\gamma_c$, planar networks undergo a buckling instability,
and wrinkled network configurations become
energetically preferable to planar configurations.
An example of a wrinkled random network is shown in
Fig.\ \ref{fig:wrex}.  We find
wrinkles at a $45^\circ$ angle to the shear direction, 
for both triangular and random networks, i.e., 
independent of microstructure. 
For both network types,  
 patterns with $n$ wrinkles are well described by a 
displacement field $z = A \sin\left(\pi y/L_y \right)
     \sin\left(\sqrt{2}\pi (y-x)/\lambda_n  \right)$ 
normal to the $xy$ plane with an amplitude $A$ and 
 wavelengths $\lambda_n =  L_x/n\sqrt{2}$, 
 as predicted   by membrane
elasticity theory \cite{Wong2006,EPAPS} ($n=3$ in Fig.\ \ref{fig:wrex}).
We focus on densities $\eta \gtrsim 30$ because wrinkling patterns are 
not well defined  in regions of  
low network density.

Continuum elasticity theory predicts a critical value
 $\gamma_c \propto B/G \propto l_b^2$ for wrinkling of 2D membranes
\cite{timoshenko, EPAPS}:
Wrinkling relieves compressional stress in the direction perpendicular 
to the wrinkles, thus, lowering the in-plane elastic energy per area to 
$e_{2D,wr} = G\gamma^2/2 -(\pi^2/2) G \gamma A^2/\lambda^2 + {\cal O}(A^4)$. 
On the other hand, 
wrinkling costs an additional bending energy 
$e_{B,wr} = 4\pi^4 BA^2/2 [\lambda^{-4}+ \lambda^{-2}L_y^{-2} + L_y^{-4}/16]$
per area.
Wrinkling sets in if $\Delta e =  e_{2D,wr}+e_{B,wr}-G\gamma^2/2 <0$ for 
the largest possible wavelength $\lambda_1 = L_x/\sqrt{2}$. 
This happens for
\begin{equation}
 \gamma >\gamma_c = (4\pi^2 B/GL_y^2)[2a^2+1+a^{-2}/32] \ ,
 \label{eq:gammac}
\end{equation}
where $a\equiv L_y/L_x$ is the aspect ratio. 
This energy argument also predicts a supercritical pitchfork 
bifurcation with $A^2 \propto \gamma-\gamma_c$  at the onset of wrinkling,
which we use to obtain $\gamma_c$ in simulations by extrapolating $A^2$ 
to zero as a function of shear angle.

 In our simulations of
 regular networks under shear, we confirm these results and find
 good agreement with (\ref{eq:gammac}), see Fig.\ \ref{fig:gcEhy}(a).
 Fitting the data with a function 
 $\gamma_c \propto l_b^{\alpha}\propto (\kappa / \mu)^{\alpha/2}$ yields
    exponents $\alpha = 1.93,...,1.97$ depending on network density.  
For random  networks, the situation 
is more complicated because we have to average over many realizations
with different $\gamma_c$.
For the average value, we find a slightly smaller exponent $\alpha = 1.9$
for $\eta = 47$.
Consequently, $\gamma_c$ follows a power law as a function 
of $(\kappa / \mu)$ with an exponent close to $1$
in semiflexible networks as predicted by continuum elasticity theory; 
the dependence on network density is weak. 
Thus, it is a generic feature  of both  regular and disordered 2D
 semiflexible polymer networks to become unstable with respect 
to wrinkling already at small critical shear angles. Therefore, 
wrinkle formation will be  relevant in most applications.

The above results were obtained using perturbations $z$ corresponding
 to a sinusoidal displacement field with wavelength $\lambda_1$.
 When applying random perturbations in the $z$ direction,
 wrinkling sets in at larger shear angles $\gamma_{c,r}$ in both
 regular and random
 networks. The difference can be as large as two orders of magnitude.
 Additionally, we observe a hysteretic behavior at the onset of
 wrinkling in random networks as shown 
 in Fig.\ \ref{fig:gcEhy}(b) (red trajectory) for   the total energy 
ratio $E_{wr}/E_{pl}$ of wrinkled and corresponding planar networks. 
Shearing beyond $\gamma_{c,r}$ and then relaxing
 the wrinkled state to $\gamma<\gamma_{c,r}$, we find that wrinkles
 persist as we would expect since $\gamma_c < \gamma_{c,r}$. 
Repeated simulations show that the threshold $\gamma_{c,r}$ for wrinkling 
decreases  for increased  amplitudes of the random 
perturbation of the unwrinkled configuration.
 This effect can 
be explained by an energy barrier which 
exists between two metastable minima in
the energy landscape,
corresponding to the unwrinkled and wrinkled states
(and exists for all $\gamma>\gamma_c$).
This barrier  is more likely to be overcome 
 for increased  noise amplitudes.

Both planar and wrinkled 2D 
networks avoid compressive stress by bending. In planar networks, the
individual segments  bend in plane
on the small scale $l_c$ between cross-links, 
 whereas wrinkling
allows bending on larger length scales as 
cross-links move out of plane. 
As a consequence, to avoid the
same amount of compressive stress, less bending energy is 
required in wrinkled networks, which gives rise to a lower 
total energy $E_{wr}<E_{pl}$. 
Figure \ref{fig:gcEhy}(c) shows that 
the maximum relative energy gain by wrinkling 
 is reached for deformations just  above the threshold
$\gamma_{c,r}$ and ranges from $10\%$ to $20\%$
in random networks (regular networks behave qualitatively 
similar). 
At higher strains, the energy gain decreases again and $E_{wr}/E_{pl}$
approaches unity for large $\gamma$.
This is a characteristic feature  of the nonlinear elasticity 
of semiflexible polymer networks 
with strain hardening  and an elasticity dominated by 
the stretching of fibers along the principal strain axis at a 
$45^\circ$ angle for large $\gamma$ \cite{Onck2005}:
Whether the material wrinkles or only bends in plane in the 
direction perpendicular to the principal strain axis has negligible 
effects on the elastic energy. 
In a  material following 
 linear continuum elasticity theory of wrinkling, we 
find  $E_{wr}/E_{pl}\approx (1+\nu)/2$ for large 
$\gamma$ instead \cite{EPAPS}.
Figure \ref{fig:gcEhy}(c)  also shows that 
the wrinkling energy gain is larger for dense networks.
 An energy gain of up to $20\%$ by wrinkling 
shows that wrinkling will be  relevant for 
practically all elastic properties and that
constraining sheetlike materials in simulations to 
planar configurations will modify material properties 
considerably.

In random networks,  wrinkles also exhibit metastability. 
This is investigated using ``preconditioned networks'', where 
we  transfer the configuration of a regular network with identical elastic 
properties to a disordered network as the initial condition for further 
energy minimization. This results in lower  total energies.
The  preconditioned networks remain energetically
favorable also at higher strains, see Fig.\ \ref{fig:gcEhy}(b) (blue
trajectories),
indicating the existence of several local minima in the energy landscape.
These correspond to different metastable configurations with 
 slightly different wrinkle configurations. 
Since the configuration with the lowest total energy is similar 
in regular and disordered networks, we also conclude that the 
wrinkling wavelength is
independent of the network microstructure.

\paragraph{Wrinkle properties.}

\begin{figure}
\begin{center}
\includegraphics[width=\linewidth]{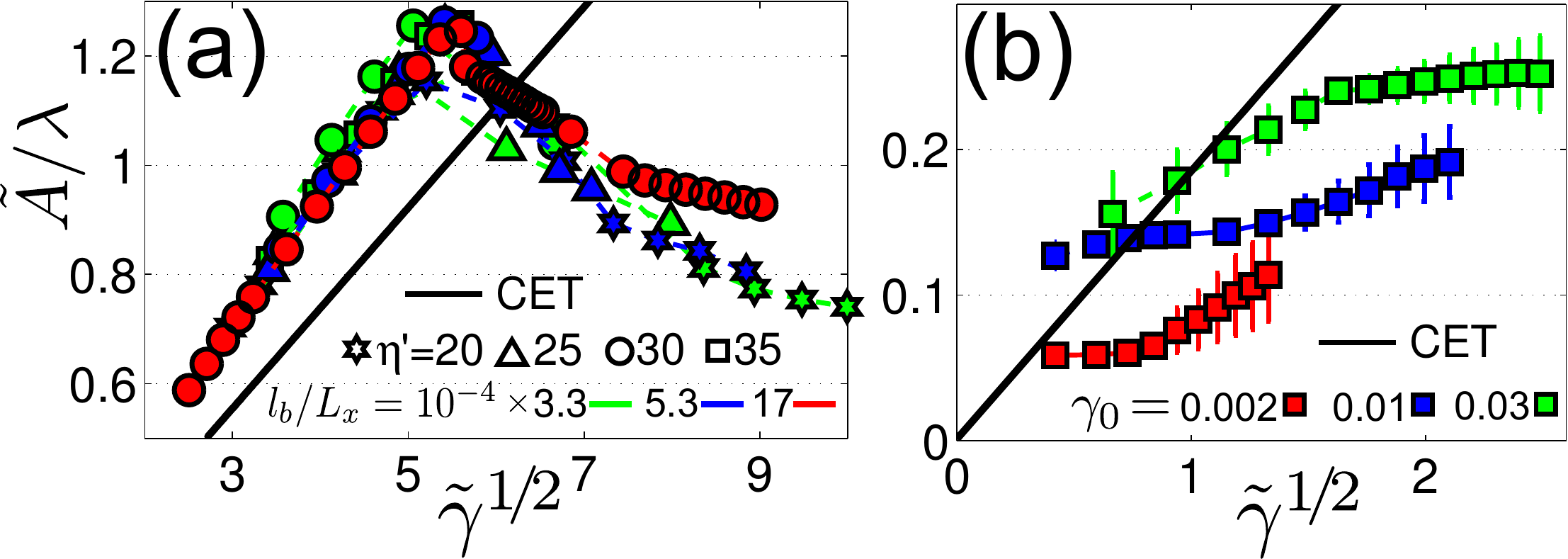}
\caption{
  Ratio of   rescaled wrinkle 
   amplitude and wavelength $\tilde{A}/\lambda$ as a
  function of $\tilde{\gamma}^{1/2}$ (a) for regular networks 
   ($l_b/L = 3.3,...,17\times 10^{-4}$ and 
   $\eta'=20,\,25,\,30,\,35$) and (b) for preconditioned random networks 
   ($l_b/L = 2.5\times 10^{-3}$ and 
   $\eta=39$).
  For regular networks (a), 
  the rescaled quantities exhibit data collapse. 
  Agreement with linear continuum elasticity theory (CET)
  [black line, Eq.\ (\ref{eq:amplitude})]
   only holds up to a maximal strain 
   $\tilde{\gamma}_{max}$.
  For preconditioned random networks (b),  rescaling does not 
  lead to  data collapse, and  the
    amplitudes  depend on the strain
     $\gamma_0$ at which the networks were preconditioned.
}
\label{fig:amps}
\end{center}
\end{figure}

Wrinkles can be characterized by their wavelength  and their
amplitude. 
In elasticity theory of  membranes, the wrinkle wavelength  $\lambda$
can be related to the applied shear. Far from the wrinkling threshold,
$\gamma\gg \gamma_c$,  the  force-equilibrium in the $z$-direction
between 
the outward bending forces $\propto BA\lambda^{-4}$ 
and the inward stretching forces $\propto Y\gamma A\lambda^{-2}$ 
gives \cite{Wong2006,EPAPS}
\begin{equation}
   \lambda =  (8\pi L_y)^{1/2}
           (B/Y)^{1/4} \gamma^{-1/4}
      = (\sqrt{72}\pi L_y l_b)^{1/2} \gamma^{-1/4}
\label{eq:lambda}
\end{equation}
where we used $B/Y =  9l_b^2/8$.
With periodic boundary conditions, $\lambda$
cannot change continuously but is restricted 
to discrete values $\lambda_n = L_x/n\sqrt{2}$. 
We find that  regular networks follow Eq.\
(\ref{eq:lambda}) from elasticity theory within the limits 
of this discretization.
For random networks, we  find that preconditioning 
with the wavelength of regular networks  always
 lowers the total energy, see  Fig.\ \ref{fig:gcEhy}(b). 
This lets us conclude  that the stable 
 wavelengths of wrinkles in random and regular
networks are identical, 
implying validity of Eq.\ (\ref{eq:lambda}) also  for random networks. 
Therefore, by measuring the wavelength of  wrinkles, 
the ratio of bending to Young's/shear modulus 
 $B/3G = 8B/9Y = l_b^2$ or local strains $\gamma$ 
can be determined via Eq.\ (\ref{eq:lambda}). 
This works  independently
 of the microstructure of the network.

A second characteristic of wrinkles is their amplitude $A$. 
In continuum elasticity of membranes, 
$A$ is a function of the wavelength $\lambda$ and
the applied shear strain $\gamma$ \cite{Wong2006,EPAPS}, 
\begin{equation}
A =  \lambda (2\pi)^{-1}{\sqrt{2(1-\nu)\gamma}},
\label{eq:amplitude}
\end{equation}
which is valid far from the wrinkling threshold. 
This result is obtained from linear continuum elasticity theory 
by assuming that wrinkling leads to a vanishing shear 
stress in the direction perpendicular to the wrinkles \cite{EPAPS}.
For small strains $\gamma$,  
our simulation results for wrinkled regular networks
show good agreement with  (\ref{eq:amplitude}) with 
  slightly larger amplitudes, see Fig.\ \ref{fig:amps}(a), 
which can be attributed to nonlinear effects. 
We find that  
the dependence $A \sim \gamma^{1/2}$ only holds 
up to a maximum strain $\gamma_{max}$
 which depends on network density and  bending stiffness, 
see Fig.\ \ref{fig:amps}(a).
 For $\gamma_{max}$, we find the empirical relation
\begin{equation}
  \gamma_{max} \sim \xi
  \equiv l_b^2 l_c^{-2} \sqrt{\lambda L_y^{-1}} \ .
\label{eq:strmax}
\end{equation}
Using $\gamma_{max} \sim \xi$  to rescale
strains to
$\tilde{\gamma} = \gamma \xi^{-1}$
and amplitudes to $\tilde{A} = A \xi^{-1/2}$,
we  achieve data collapse onto a master curve 
with a maximum reduced amplitude at 
$\sqrt{\tilde{\gamma}_{max}} \approx 5.4$.
Typical values for $\xi$  are $\xi = 2.9,...,68\times 10^{-5}$.
Only for values $\tilde{\gamma}
< \tilde{\gamma}_{max}$, the  wrinkle amplitude is in 
agreement with Eq.\ (\ref{eq:amplitude}). Hence, $\gamma_{max}$ gives 
an estimate for the maximum strain at which it is reasonable to
treat regular networks and their wrinkles by  elasticity theory.
 According to Eq.\ (\ref{eq:lambda}),
the corresponding wavelength $\lambda(\gamma_{max})$
only depends on $l_c$, which suggests
that for $\gamma >\gamma_{max}$, the discrete network structure 
becomes relevant.

 When comparing random networks to regular ones, we notice
   deviations from Eq.\ (\ref{eq:amplitude}) already at smaller
  strains. Moreover, after being
   preconditioned at different strains, random  networks exhibit a pronounced 
    dependence on the initial
   state resulting in  significantly different  amplitudes  as shown in 
   Fig.\ \ref{fig:amps}(b). This indicates 
  the existence of different  metastable
   configurations differing in amplitudes despite
   having the same wavelength.

\paragraph{Crossover from bending to stretching.}

Typically, the nature of the deformation in  planar random   
networks depends on the strain and network density. 
At small strains, linear elastic properties are bending 
dominated for small densities and become stretching dominated 
for higher densities \cite{Head2003,Wilhelm2003,Head2003b,Heussinger2006}.
For a fixed density, networks are  bending dominated at low strains 
and become stretching dominated at high strains, which 
results in  nonaffine deformations at the 
crossover \cite{Onck2005}.
The ratio of bending to stretching energy correlates
with measures for the nonaffinity of network deformation and can, thus, 
be used as an indicator for the transition in deformation behavior 
  in wrinkled networks. 
Comparing the same networks in planar and three-dimensional simulations, we
 find that wrinkling causes a transition to stretching dominated behavior 
 at much smaller strains than in planar networks. 
Our results suggest that the wrinkling transition inherently causes the
 network to become stretching dominated even if the strain is too small to
 cause stretching domination in a planar network. 
Details are included
 in the Supplemental Material \cite{EPAPS}.

\paragraph{Conclusion.}

We investigated wrinkling of two-dimensional 
random and triangular semiflexible 
polymer networks under shear. 
Both types of  networks 
wrinkle already  above  small  critical shear angles with 
$\gamma_c \propto (\kappa / \mu)^{\alpha / 2}$, 
where $\alpha$ ranges from $1.9$ to $2.0$
[see Fig.\ \ref{fig:gcEhy}(a)]. 
Random networks exhibit hysteresis near the wrinkling
 threshold
[see Fig.\ \ref{fig:gcEhy}(b)]. 
The maximal energy gain upon wrinkling is 
up to $20\%$. Therefore, wrinkling is a relevant effect for 
all elastic properties [see Fig.\ \ref{fig:gcEhy}(c)].
In random networks, we found metastable 
wrinkled configurations.
Even though the  disordered microstructure in random networks
strongly affects the 
scaling properties of the wrinkle amplitude [see Figs.\ \ref{fig:amps}(a) and 
   \ref{fig:amps}(b)],
it has little effect on the wrinkle wavelength, which  follows predictions 
from elasticity theory.
Therefore, wrinkle wavelengths 
can represent a robust, microstructure-independent 
assay of shear strain or elastic properties.
Wrinkling strongly affects the characteristic 
elastic response of 2D semiflexible polymer networks
since it  triggers an immediate transition to 
stretching dominated deformation.

\paragraph{Acknowledgments.}
 We acknowledge financial support by the 
 Mercator Research Center Ruhr (MERCUR).


\end{document}